\documentclass[12pt]{iopart}
\usepackage{graphicx}  
\begin{document}

\title[Lithographic MBCJs for single-molecule measurements in vacuum]{Lithographic mechanical break junctions for single-molecule measurements in vacuum: possibilities and limitations}

\author{Christian A Martin$^{1,2}$, Dapeng Ding$^{1,2}$, Herre S J van der Zant$^1$ and Jan M van Ruitenbeek$^2$}
\address{$^1$ Kavli Institute of Nanoscience, Delft University of Technology, Lorentzweg 1, 2628 CJ Delft, The Netherlands}
\address{$^2$ Kamerlingh Onnes Laboratorium, Universiteit Leiden, Postbus 9504, 2300 RA Leiden, The Netherlands}
\ead{ruitenbeek@physics.leidenuniv.nl}

\begin{abstract}
We have investigated electrical transport through the molecular model systems benzenedithiol, benzenediamine, hexanedithiol and hexanediamine. Conductance histograms under different experimental conditions indicate that measurements using mechanically controllable break junctions in vacuum are limited by the surface density of molecules at the contact. Hexanedithiol histograms typically exhibit a broad peak around 7$\cdot 10^{-4}$~$G_0$. In contrast to recent results on scanning tunneling microscope (STM)-based break junctions in solution we find that the spread in single-molecule conductance is not reduced by amino anchoring groups. Histograms of hexanediamine exhibit a wide peak around 4$\cdot 10^{-4}$~$G_0$. For both benzenedithiol and benzenediamine we observe a large variability in low-bias conductance. We attribute these features to the slow breaking of the lithographic mechanically controllable break junctions and the absence of a solvent that may enable molecular readsorption after bond breaking. Nevertheless, we have been able to acquire reproducible current-voltage characteristics of benzenediamine and benzenedithiol using a statistical measurement approach. Benzenedithiol measurements yield a conductance gap of about 0.9~V at room temperature and 0.6~V at 77~K. In contrast, the current-voltage characteristics of benzenediamine-junctions typically display conductance gaps of about 0.9~V at both temperatures.
\end{abstract}

\textit{for final version see: New Journal of Physics 10 (2008), 065008}.

\section{Introduction}
Since the first measurements of the current-voltage ($IV$) characteristics of a benzenedithiol molecule~\cite{reed97}, research in single-molecule electronics has progressed rapidly. Today, numerous techniques based on nanolithography and scanning probe instruments allow for studies of the electrical properties of single molecules~\cite{ruitenbeek96,park99,li98_2,xu03,haiss03,donhauser01,cui01}. Nevertheless, the field of molecular electronics is far from maturity. The influence of chemical structure and anchoring on the electrical characteristics of single molecules remains a fascinating and largely unexplored area of research.\\

The low-bias conductance of organic molecules is mostly investigated with the break junction method. Here, a metallic contact is repeatedly broken and re-established in the presence of the molecules of interest. This principle builds on the assumption that molecules binding to the metal surfaces will, with some finite probability, form a conducting bridge between both electrodes after cleaving the last metal-to-metal contact. Stable conductance values during the stretching of the junctions are hence attributed to single-molecule contacts.\\
To date, most break junction measurements have been carried out in solution using scanning probe instruments. Although this appears to be a very suitable environment for studying single-molecule properties, it also imposes an important limit on the experiments: The temperature range is restricted to a few tens of degrees.\\
Mechanically controllable break junctions (MCBJs) do not suffer from this restriction. They can easily be operated down to cryogenic temperatures in a vacuum environment. Furthermore, the electrode separation in these devices exhibits exceptional stability across the entire temperature range~\cite{ruitenbeek96}. This is mainly due to the working principle of MCBJs, in which the bending of a flexible substrate is translated into the stretching of a metal wire on the device surface. The ratio of these displacements can be as low as $10^{-5}$ for lithographic devices, giving rise to sub-\AA ngstrom control of the electrode separation~\cite{vrouwe05}. However, this so-called displacement ratio also limits the measurement speed of lithographic MCBJs.
They are usually broken at not more than 1~nm/s~\cite{gonzalez06,loertscher07}. STM break junctions, in contrast, can be operated at maximum breaking speeds of more than 100~nm/s.\\
This seemingly technical difference can have large consequences for break junction experiments. The importance of stretching speed for the lifetime of single-molecule junctions has only recently been appreciated~\cite{tsutsui08,huang07}. Experimental results suggest that the breaking mechanism of metal-molecule bonds changes from a force-dominated one at fast stretching to spontaneous breakdown at low speeds. Furthermore, the chemical nature of the bond is of large importance for the lifetime of single-molecule junctions~\cite{huang07,park07}. Accordingly, the statistical weight of molecular conductance steps as compared to the experimental signature of vacuum tunnelling will strongly depend on both the breaking speed of the junction and the molecular anchoring group.\\

In break junction experiments at room temperature, molecular contacts are mostly formed in a solution of the species of interest. For low-temperature measurements, however, one has to apply different schemes of molecule deposition. Small molecules like hydrogen can be condensed from the vapour phase~\cite{smit02}. In contrast, the application of larger molecules usually requires a solution phase before sample mounting and cool-down. Common approaches for molecule deposition range from the drop-casting of dilute solutions on open junctions~\cite{reed97,loertscher07,kergueris99,reichert02} to the formation of a (sub-)monolayer with subsequent rinsing~\cite{fujihira06,li07_2}. The latter method avoids complications due to the crystallization and oligomerization~\cite{tour95} of thiols on the open junction. 
We have adapted this method to experiments with lithographic MCBJs using benzene and hexane with thiol and amino end groups.\\

Alkanedithiols are prototype systems for the study of single-molecule electronic properties. This is due to their comparatively simple chemical structure. Their anchoring to metals is achieved through thiol groups (SH), which have been studied since the earliest stage of molecular electronics~\cite{tour95}. This is largely due to the well-developed chemistry of thiol self-assembly, which guarantees a dense coverage of both flat and rough gold surfaces~\cite{love05,losic01,kawasaki00}. Thiols are expected to bind to three sites on the Au(111) surface, the so-called top, hollow and bridge site~\cite{fujihira06}. Metal point contacts, which exhibit a larger roughness, might even allow for more degrees of freedom in binding and conformation~\cite{love05}. The chemical versatility of the gold-sulfur bond can, in fact, have detrimental consequences for their electronic properties. Simulations suggest that individual configurations exhibit varying electronic properties both in terms of the electronic coupling and the Fermi-level alignment~\cite{basch05,tomfohr04,muller06}. Experimental observations on a molecule as simple as hexanedithiol support this result. Literature data display a spread over more than two orders of magnitude in conductance and range from 2$\cdot 10^{-5}$ to 2$\cdot 10^{-3}$~$G_0$~\cite{xu03,haiss04,nishikawa07,jang06,ulrich06,chen06}. Here, $G_0={2e^2}/{h}$, the quantum of conductance. Likewise, studies of the prototypical $\pi$-conjugated benzenedithiol have yielded low-bias conductances between 5$\cdot 10^{-5}$ and 1$\cdot 10^{-1}$~$G_0$~\cite{loertscher07,ulrich06,tsutsui06,he05}.\\
Only recently, a succesful attempt to reduce this ambiguity has been reported. By changing from thiol to amino (NH$_2$) anchoring groups, Venkataraman~\etal~\cite{venkataraman06} were able to reduce the spread of molecular conductance in STM-based break junctions in solution considerably. Amines have been used only scarcely in the functionalization of metal surfaces, and their bond to gold is weaker than that of thiols~\cite{xu93,brown99,kumar03}. Simulations suggest that the ability to extract single gold atoms from solid surfaces, which has been reported for thiols, is not present in amines~\cite{li07_1}. 
Their well-defined conductance may be due to the preferential binding to surface adatoms. This motif not only limits the possible adsorption geometries, but its coupling is also relatively independent of bond rotation and tilt angle~\cite{li07_1,quek07}.\\

We have tested the applicability of the thiol and amino anchoring scheme in lithographic mechanically controllable break junctions in vacuum. The model systems used in these experiments were 1,6-hexanedithiol (HDT), 1,6-hexanediamine (HDA), 1,4-benzenedithiol (BDT), and 1,4-benzenediamine (BDA). Finally, we have measured $IV$ characteristics for the two benzene derivatives. Whilst two measurements of stable $IV$ characteristics of BDT have been reported~\cite{reed97,loertscher07}, we do not know of any such experiments on BDA to date.

\section{Experimental details}
\subsection{Fabrication of MCBJs}
Phosphorous bronze wafers (50~mm x 50~mm x 0.3~mm) were polished and cleaned by ultrasonication in acetone and isopropanol (IPA). After the application of an adhesion promoter (VM651, HD Microsystems) the wafers were spin-coated with a commercial polyimide precursor solution (PI2610, HD Microsystems). We cured the film in a vacuum oven at a maximum temperature of 300$^\circ$C in order to fully crosslink the precursor and expel remaining solvent. The thickness of the resulting polyimide layer was around 3~$\mu$m, sufficient for electrical isolation of the break junction devices from the metallic substrate.\\
For the subsequent electron-beam lithography step the wafers were spin-coated with a double layer of resist (320~nm of (methylmethacrylate (8.5) methacrylic acid) copolymer, followed by 110~nm of polymethylmethacrylate (PMMA) 950k). The resists were spin-coated from ethyl-L-lactate and anisole, respectively (Microchem). Both layers were baked out for 15~min at 175$^\circ$C. We then defined 10 devices on a wafer with 4 break junctions in each device using a Leica electron-beam pattern generator. In each junction two contact wires narrow down into a thin bridge, which is 100~nm wide and 200~nm long. Lithography was carried out at an acceleration voltage of 100~kV. After exposure we developed the patterns in a mixture of methyl-isobutyl-ketone and IPA (1:3) for 90~s. We then deposited 1~nm of Cr and 80~nm of Au in an electron beam evaporator at a base pressure of $10^{-7}$ mbar. The evaporation rates were 0.5~\AA/s and 1.0~\AA/s, respectively. After lift-off in hot acetone and a rinsing step in IPA we protected the fully processed wafers with a 500~nm thick layer of PMMA 350k. The individual break junction devices were obtained later by laser-cutting and removing the protection layer (see below).

\subsection{Preparation of molecular solutions}
Shortly before each measurement we prepared a fresh solution of the molecules of interest in toluene. All glassware used during the preparation procedure was initially cleaned by ultrasonication in fuming nitric acid, followed by ultrasonication in deionized water. After drying, the glassware was ultrasonicated once more in pure toluene (high performance liquid cromatography (HPLC) grade, Aldrich). We then bubbled toluene with nitrogen for at least 30~min in order to expel dissolved oxygen from the solvent and increase the reproducibility of the self-assembly process~\cite{love05}. BDA, HDT, HDA were used as received from Aldrich. BDT was obtained from Tokyo Chemical Industry (TCI). The concentration of all solutions was around 1~mmol/l.

\subsection{Sample preparation}
In the first step of the preparation, we removed the protection layer of the individual devices by immersion in hot acetone, followed by immersion in hot IPA. The devices were then etched in a Leybold RF plasma etcher using a mixture of CF$_4$ and O$_2$ gas (flow rate ratio 1:4, pressure 0.3~mbar, RF power 20~W). This results in a nearly isotropic etch profile with a suspended electrode length of about 2~$\mu$m (see figure~1). Immediately after etching we dipped the devices into fresh ethanol (Baker, 99.9\%) to reduce possible gold oxide on the surface~\cite{ron98}. Gold oxide can be encapsulated under alkanethiol monolayers and reduce the reproducibility of self-assembly processes~\cite{ron98,ron94}.
\begin{figure}
\label{fig_mcbjs}
\centering
\includegraphics{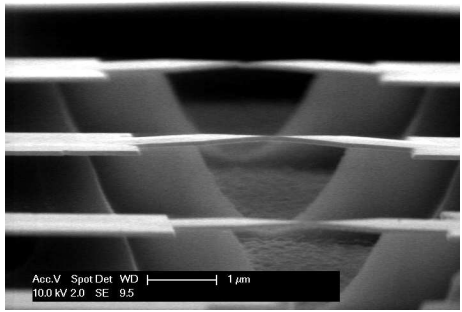}
\caption{Scanning electron micrograph of a typical lithographic MCBJ sample. The micrograph displays three out of a total of four junctions, which break independently during bending.}
\end{figure}
Immediately before the controlled monolayer formation each sample was rinsed in pure toluene and blown dry in a flow of nitrogen gas. We then placed a small drop of the molecular solution on the area of the junctions and left the devices in a toluene-saturated atmosphere for 1 minute. After self-assembly, the devices were rinsed thoroughly in toluene and blown dry with nitrogen.

\subsection{Measurements}
We mounted the MCBJs into a home-made vacuum insert equipped with a mechanical feedthrough. Before measurement, the setup was evacuated to a base pressure of $10^{-5}$~mbar. A brushless DC servo motor was used to actuate the three-point bending mechanism of the MCBJ, which was based on a differential screw with a pitch of 250~$\mu$m per turn.\\
All MCBJ measurements were automated using the Labview software package. We measured low-bias conductance histograms at a DC bias of 50~mV. In order to break and re-establish the nanoscale contact (further referred to as breaking and making) the flexible MCBJ substrates were bent at an actuator speed of 2~$\mu$m/s. We used the conductance as a control parameter for this process. Thus, the junctions were cycled between the open and the fused state, given by conductances below 1$\cdot 10^{-5}$ and above a few $G_0$, respectively. Real-time analysis of the conductance allowed for motion reversal in less than a millisecond. In order to ensure complete opening of the junction the substrate was automatically deflected by another 50~$\mu$m after the conductance had crossed the lower limit. A home-built logarithmic $IV$-converter with a temperature-drift compensation was used to measure the current during bending, covering a range from less than 50~pA to more than 1~mA. Pointwise calibration with standard resistors ranging from 100~Ohm to 1~GOhm ensured the accuracy of this conversion, similar to experiments reported by He~\etal~\cite{he05}.\\
We carried out the $IV$ measurements on benzene derivatives in a statistical scheme similar to the one introduced by L{\"o}rtscher~\etal~\cite{loertscher07}. The MCBJs were bent in discrete steps of 0.5 to 0.7~$\mu$m. At each position, we first determined the low-bias conductance at 25~mV and then acquired 3 to 4 $IV$ traces. We chose to measure the $IV$ characteristics in the bias interval from -1.2 to 1.2~V, starting from 0~V. The junction currents were measured using a linear $IV$ converter at $10^6$~V/A, which resulted in sub-nA resolution. Furthermore, this setting limited the maximum current to 4~$\mu$A. During the systematic $IV$ measurements the low-bias conductance at 25~mV was used as a control parameter for the substrate bending. Thus, the maximum junction conductance during making was limited to 2.1~$G_0$.

\section{Results and discussion}
\subsection{Optimization of measurement parameters}
Typical breaking traces of lithographic MCBJs in vacuum are presented in figure~2(a). As the junctions are stretched, their conductance decreases from several $G_0$ to a value below the noise floor of our setup. During this process, the conductance does not evolve continuously, but it exhibits sudden jumps and plateaus.
In junctions exposed to clean toluene the last pronounced plateau repeatedly occurs around 1~$G_0$. It is well established that this value corresponds to a single gold atom bridging the gap between the electrodes~\cite{agrait03}. Upon further stretching, the conductance of the clean junctions decreases roughly exponentially with occasional jumps and fluctuations.
In Au MCBJs, the observation of a plateau at the quantum of conductance is relatively robust and independent of measurement parameters. It is also present in devices that have been exposed to molecular solutions. In addition, many of the breaking traces on these junctions display steps of varying length and height below 1~$G_0$. The conductance at these plateaus can exhibit significant fluctuations both at room temperature and at 77~K.\\

Due to microscopic variations in the contact geometry and the evolution of the conductance a large number of breaking traces has to be acquired and analyzed statistically. This can be achieved by means of conductance histograms~\cite{krans93}. Individually sampled data points from trace measurements serve as the smallest possible elements in the construction of these histograms. The entire conductance range is split into short intervals, so-called conductance bins. Each time a particular conductance value is recorded the number of counts in the respective bin is increased. In the final histogram, the largest number of counts indicates the most stable junction configuration.\\
We constructed histograms from 300 subsequent breaking traces of individual junctions without any further data selection. We would like to note that this approach differs from most other experiments. Often, only traces displaying plateaus are used to construct histograms. We have chosen to include all breaking data for two reasons. First, we are interested in the influence of measurement parameters on the behaviour of the junctions - a selection of traces might prevent us from drawing valid conclusions. Secondly, we think that conductance noise and jumps are typical of molecular junctions. Traces exhibiting these features should be included in the statistical analysis.\\
We analyzed all breaking data using logarithmically spaced conductance bins~\cite{gonzalez06}. This allows us to display the junction conductance over a much wider range than in the widely established linear representation. In addition, conductance plateaus lead to sharper features when using logarithmic histograms. We normalized the counts in all bins to an equivalent of milliseconds per trace in order to prevent sampling rates from affecting the histograms. This allows for comparisons across different experimental platforms and illustrates the time scales of the breaking process.\\

The breaking histogram of a typical junction exposed to pure toluene is presented in figure~2(b).
\begin{figure}
\label{fig_conditions}
\centering
\includegraphics{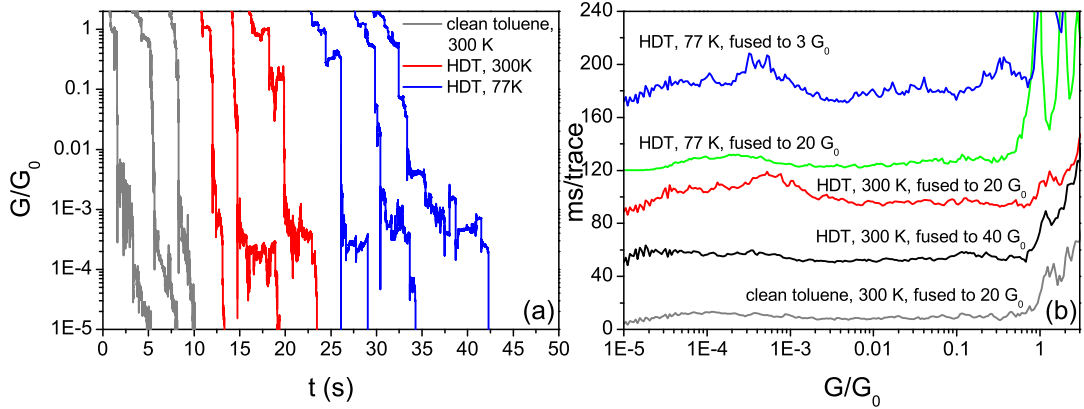}
\caption{Breaking characteristics of MCBJs exposed to clean toluene and to solutions of hexanedithiol. The junctions were broken at a bias of 50~mV. (a) Typical breaking traces. (b) Conductance histograms for different temperatures and fusing depths. Each histogram was constructed from 300 subsequent breaking curves of one junction using 42 bins per decade. Curves are offset by 40~ms/trace for clarity.}
\end{figure}
In contrast to previous results on lithographic MCBJs in solution~\cite{gonzalez06}, we systematically observe detectable counts across the entire range below 1~$G_0$. The counts are distributed rather uniformly, so that there are no indications for preferred conductance values. This uniform distribution can be rationalized by investigating the breaking traces. When compared to other data, particularly from STM measurements~\cite{ulrich06,venkataraman06}, the traces measured in our MCBJs exhibit a much less pronounced conductance drop after the 1~$G_0$ plateau. Besides, we repeatedly record fluctuations in the regime of vacuum tunnelling.\\

We believe that impurities from the solvent can be excluded as a possible cause of this difference. First, control experiments on samples not exposed to any solvent did not lead to a reduction of counts below 1~$G_0$. Furthermore, we fused devices exposed to toluene to conductances of more than 40~$G_0$ to create fresh breaking surfaces. Also in this case, the background signal in the tunnelling regime could not be reduced.\\
We think that the uniform distribution of histogram counts below 1~$G_0$ must rather be related to the small indentation and the slow breaking of our lithographic MCBJs. Both may have an influence on the micromechanics of their contact area and their breaking behaviour. The speed of contact separation in our junctions is on the order of 60~pm/s, which is by a factor of 20 slower than in the experiment reported by Gonz{\'a}lez~\etal~\cite{gonzalez06}, and by more than 2 orders of magnitude below typical breaking speeds in STM~\cite{huang07}.\\
During fast breaking of Au MCBJs, the contact separation usually increases spontaneously after cleaving the last gold-gold bond. This can either be due to the relaxation of atomic chains~\cite{yanson98} or due to the release of mechanical stress in the banks~\cite{rubio96}. In histograms, such a spontaneous opening leads to a depletion in counts in the high tunnelling regime. In slow measurements at room-temperature, thermal fluctuations may reduce the probability of forming chains significantly. Furthermore, the gentle deformation of our lithographic MCBJs may prevent mechanical stress in the banks, leading to a more uniform tunneling signal in the breaking histogram. The slow breaking speed may also explain the comparatively small and broad peak at 1~$G_0$ in our measurements. Thermal fluctuations can limit the lifetime and the statistical signature of the last gold-gold bond. The observed histogram peak height of 40~ms per trace is on the order of the natural lifetime of the gold-gold point contact obtained by Huang~\etal~\cite{huang07}.\\
 
We have systematically studied the importance of the measurement parameters on histograms of the model system hexanedithiol. The fusing depth of the HDT-covered junctions in histogram measurements turned out to be a crucial parameter for the detection of molecular features. At room temperature, fusing the MCBJs to 40~$G_0$, i.e. to about 50\% of their conductance after fabrication, did not result in a significant increase of the histogram counts below 1~$G_0$ as compared to junctions exposed to clean toluene (see figure~2). A target conductance of a few $G_0$, in contrast, led to increasing difficulties in closing the junctions to high conductances after less than 100 cycles (not shown). Fusing the MCBJs to 20~$G_0$ eventually enabled us to carry out reliable histogram measurements comprising hundreds of breaking curves.\\
A histogram obtained with these parameters is presented in the centre of figure~2(b). Compared to the data on clean toluene, the HDT histogram displays a pronounced peak around 7$\cdot 10^{-4}$~$G_0$. It indicates a higher probability of measuring this conductance value in breaking traces, which is usually attributed to the formation of metal-molecule-metal junctions (for a more detailed discussion see the following section). At 77~K, in contrast, fusing the MCBJs to 20~$G_0$ completely suppresses any peaks in the histograms. A target conductance of 3~$G_0$, corresponding to a few atoms only, again leads to signatures of the formation of molecular junctions, represented by a broad peak around 4$\cdot 10^{-4}$~$G_0$.\\

These observations suggest that single-molecule measurements in vacuum may be limited by the surface density and diffusion of thiols. Whilst insufficient fusing leads to a too dense coverage of the electrodes and prevents them from being fused after a number of cycles, deep indentation reduces the probability of creating single-molecule contacts almost completely. 

\subsection{Conductance histograms of model systems}
We have used the optimized parameters discussed above in histogram measurements of the molecular model systems hexanedithiol, hexanediamine, benzenedithiol and benzenediamine at room temperature. The junctions were cycled between the open state and 20~$G_0$ in order to increase the chance of trapping molecules during breaking.\\
Figure~3 presents histograms obtained on the substituted alkanes, hexanedithiol and hexanediamine.
\begin{figure}
\label{fig_hexane}
\centering
\includegraphics{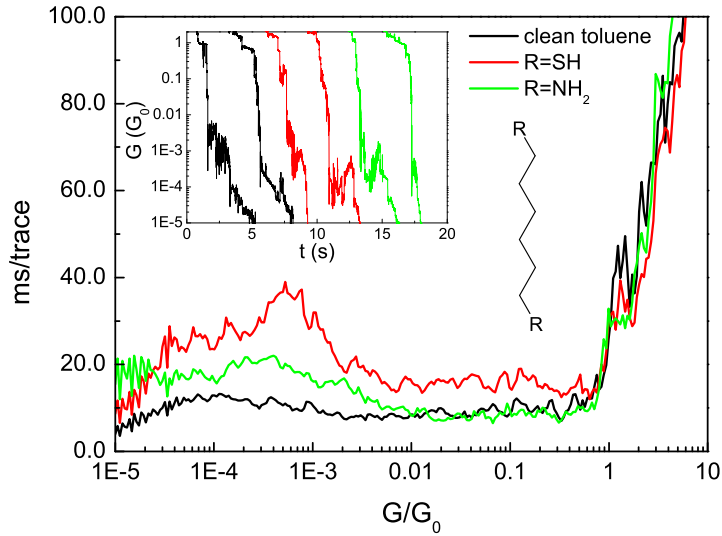}
\caption{Conductance histograms of MCBJs exposed to clean toluene and junctions containing hexanedithiol and hexanediamine. Each of the histograms was constructed from 300 subsequent breaking curves of one junction without any further data selection. The statistics were carried out using 42 bins/decade. The inset shows examples of the conductance traces measured during breaking.}
\end{figure}
Histograms obtained for HDT at room temperature displayed a broad peak around 7$\cdot 10^{-4}$~$G_0$. Depending on the experiment, peaks below this value could also be observed. The histogram in figure~3, for example, contains a series of smaller peaks around 3$\cdot 10^{-5}$, 7$\cdot 10^{-5}$, 1.5$\cdot 10^{-4}$~$G_0$.\\
The presence of more than one peak is not unusual - STM-based experiments on HDT in solution have yielded two peaks at 3$\cdot 10^{-4}$ and 1$\cdot 10^{-3}$~$G_0$~\cite{chen06}. Similar measurements in vacuum even showed 3 different conductance regimes at 3$\cdot 10^{-5}$, 2$\cdot 10^{-4}$, and 3$\cdot 10^{-3}$~$G_0$~\cite{fujihira06}. The conductance observed in our measurement falls well within this range, but there is not a close agreement with one previously published dataset.\\
Recent theoretical and experimental results shed more light on this difference. In general, the presence of several peaks can be attributed to different molecular conformations or varying contact geometries. Series of peaks corresponding to integer multiples of a basic conductance are sometimes associated with junctions containing several molecules in parallel. Furthermore, density functional theory (DFT)-based simulations and delicate STM-based experiments have indicated that different core and contact geometries of alkanedithiols may give rise to a conductance spread over 2 decades. Li~\etal attribute most of this spread to conformational changes in the alkane units~\cite{li07_2}. This effect can be of foremost importance in the case of sparse monolayers, where a large degree of steric freedom is given. The molecular concentrations and self-assembly times in our work are comparable to those of Li~\etal~\cite{li07_2}, who reported a mixture of high and low-coverage phases.\\

Surprisingly, HDA exhibits a considerably smaller and broader peak than HDT. The conductance of about 4$\cdot 10^{-4}$~$G_0$ is in reasonable agreement with previous reports on HDA, which indicate conductances of 1.2$\cdot 10^{-4}$ and 3$\cdot 10^{-4}$~$G_0$~\cite{chen06,venkataraman06}. However, the large peak width is in strong contrast to the results obtained by Venkataraman~\etal, who observed a reduction in spread of the conductance of amines with respect to thiols~\cite{venkataraman06}.\\
We believe that this difference is due to the different experimental parameters. Recent STM experiments at varying breaking speeds in solution suggest that bond re-formation after desorption can increase the lifetime of single-molecule junctions~\cite{huang07}. This effect may be suppressed in the absence of a stabilizing solvent and at a sparse surface coverage. Furthermore, simulations and experiments on amine junctions have indicated that these break faster upon stretching than thiols~\cite{park07,li07_1}. In fact, the Au-S bond is even more stable than the Au-Au bond itself, allowing for the deformation of gold electrodes during pulling~\cite{xu03_2}. In the case of a measurement limited by the lifetime of a bond, the signature of amines in histograms will be much smaller than the one of thiols. This could be crucial in lithographic MCBJs, which are broken at a very slow speed. The resulting long traces in the vacuum-tunnelling regime lead to an enhanced histogram background below 1~$G_0$, which may obscure the signature of molecular conductance steps.\\

In histograms of MCBJs exposed to benzenedithiol (see figure~4), the entire conductance regime below 1~$G_0$ displayed enhanced counts. 
\begin{figure}
\label{fig_benzene}
\centering
\includegraphics{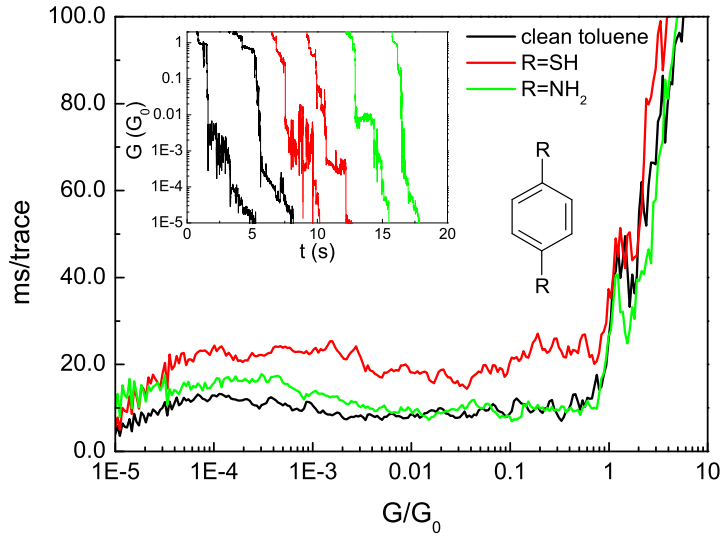}
\caption{Conductance histograms of MCBJs exposed to pure toluene and junctions containing benzenedithiol and benzenediamine. Data from 300 subsequent breaking curves were used to construct each histogram with a bin size of 42 bins/decade. The inset shows examples of breaking curves of the junctions.}
\end{figure}
The exact probability distribution also showed variation between different experiments. Above an enhanced background of counts, small peaks in the region between 0.1 and 1~$G_0$ and just above 1$\cdot 10^{-3}$~$G_0$ are visible. Overall, these are much less pronounced than the ones observed in the histograms of HDT.\\
The previously reported peak at 0.011~$G_0$~\cite{xiao04} is not evident in our data. Nevertheless, the high conductances are in the range of previous data from MCBJ measurements~\cite{tsutsui06}, and the small peak around 1$\cdot 10^{-3}$~$G_0$ agrees with recent results from STM measurements~\cite{ulrich06}. However, the broad distribution of junction conductances clearly dominates the histogram. A similar spread in conductance values has been observed by Ulrich~\etal~\cite{ulrich06}. It was attributed to geometrical variations in the metal-molecule junctions. We think that both the binding motif of the thiol groups and possible interactions between the benzene core and the gold electrodes contribute to the variation in our data. In the absence of a stabilizing solvent, even junction configurations with a molecule lying flat on the gold surface are conceivable. Due to the interaction of $\pi$-orbitals with the electrodes their electronic coupling could differ considerably from that of upright molecules, leading to different conductance values.\\

Similar to the measurements on the hexane derivatives, the use of amino anchoring groups did not lead to clearer peaks in the histogram measurements of benzenediamine (see figure~4). In its conductance histogram the region below a few $10^{-3}$~$G_0$ exhibits increased counts, but it misses a pronounced peak structure. This is different from previous data obtained in solution~\cite{venkataraman06}, but in agreement with our measurements on HDA. Again, the absence of a solvent and a fast rupture of the metal-molecule bond must have reduced the probability of forming stable molecular junctions, which are a prerequisite for pronounced peaks in the histograms.

\subsection{$IV$ measurements on benzene derivatives}
During the stepwise breaking and making of MCBJs exposed to benzenedithiol and benzenediamine we repeatedly observed series of reproducible $IV$ curves that displayed a low conductance around zero bias and a step-like increase in current at higher bias voltages. Within these series, which typically lasted for a few micrometers of bending, both the shape and the current level of the $IV$ characteristics remained similar.\\
An example of such a series of reproducible current-voltage characteristics is presented in figure~5(a).
\begin{figure}
\label{fig_IV_groups}
\centering
\includegraphics{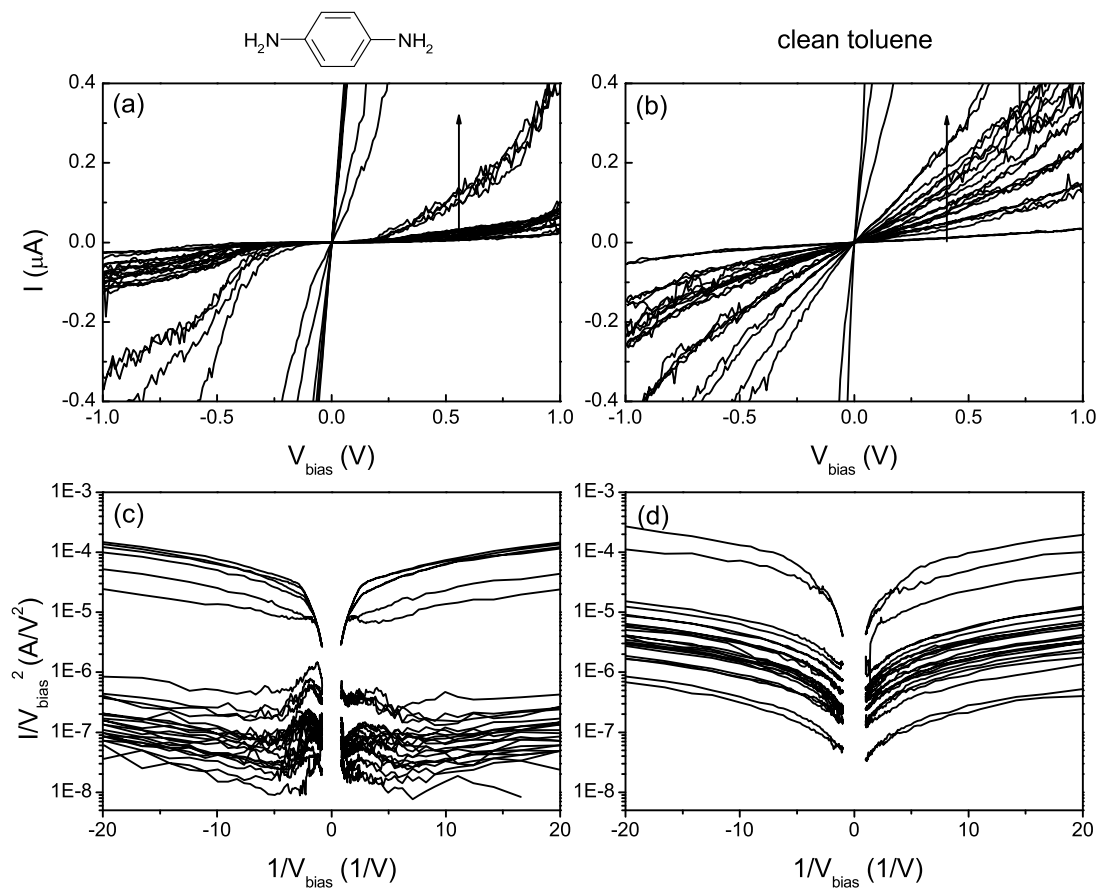}
\caption{Consecutive $IV$ traces obtained during the stepwise closing of lithographic MCBJs at room temperature. The change in electrode distance between the steps was on the order of 15~pm. The arrows mark the time-evolution of the characteristics. (a) Consecutive $IV$s obtained during the closing of a junction exposed to BDA in toluene. Groups of non-linear characteristics can be observed. (b) Smooth $IV$ curves typical of junctions exposed to clean toluene. The current increases rather steadily with decreasing gap size. (c) A Fowler-Nordheim plot of the curves for BDA. (d) A Fowler-Nordheim plot of the $IV$ characteristics for clean toluene. Both plots do not show the characteristic decrease of $I/V^2$ with $1/V$ expected for Fowler-Nordheim tunneling. Curve shapes in (d) agree with the Simmons model for tunnelling.}
\end{figure}
For a large contact separation, the junction exhibits rather smooth $IV$s with low currents. As the electrodes approach in discrete steps, the measurement yields a group of $IV$ curves that barely change upon pushing. They are marked by a distinct nonlinearity, i.e. a suppression of the junction current in the bias window from -0.3~V to 0.2~V. After several steps in the making of the junction, this series ends and a second group of non-linear characteristics is measured, again exhibiting a current suppression around zero bias. Current levels in this series are considerably larger than in the first one, and the conductance gap differs slightly. Upon moving the electrodes closer together, the gap structure is lost. Instead, smooth $IV$ curves with high current levels are measured. These high-conductance curves saturate around 4~$\mu$A (not shown in the graph), corresponding to the compliance of the measurement electronics. It is important to note that the conductance of these junctions is below 1~$G_0$ and their characteristics are slightly superlinear. Only after fusing the MCBJ further a contact with a conductance above 1~$G_0$ is formed.\\
Control experiments on MCBJs exposed to pure toluene did not show reproducible $IV$-characteristics with step-like increases in current. The low-bias conductance of these junctions evolved rather steadily in the interval from less than $5\cdot 10^{-6}$~$G_0$ to 2.1~$G_0$. An example of this evolution is given in figure~5(b). As the junction is fused, the maximum current in the smooth $IV$ characteristics slowly increases from the nA-range to about 4~$\mu$A.\\

The observation of reproducible, non-linear $IV$s during the making of MCBJs exposed to organic molecules agrees well with recent statistical $IV$ measurements by L{\"o}rtscher~\etal~\cite{loertscher07}. In their experiments, three different regimes of transport were identified: conduction through vacuum (described by smooth tunnelling characteristics with low currents), transport through a single molecule (marked by highly non-linear $IV$ characteristics), and a fused metal-metal connection (with an ohmic bias response). The transition between these regimes was very abrupt, with signatures of single-molecule junctions appearing exclusively during the making of the MCBJ. In our MCBJs, in contrast, reproducible, non-linear $IV$s could be measured both during breaking and making. Furthermore, the transition between the different regimes was less systematic and less pronounced.\\
Both differences may be related to our method of molecule deposition. Whilst L{\"o}rtscher~\etal~\cite{loertscher07} applied BDT to open junctions, we carried out the self-assembly before breaking. This may lead to different adsorption geometries and surface densities of the molecular (sub-)monolayer in the contact area, and possibly a larger variation in the probed metal-molecule-metal junctions.\\
It is important to note that the $IV$ characteristics of junctions exposed to BDT and BDA were not always symmetric. Generally, the presence of asymmetric $IV$s agrees with previous reports on single-molecule measurements~\cite{loertscher07,reichert02}. Within the series of reproducible $IV$ characteristics the asymmetry did not change significantly, nor did the overall current level. Therefore, the asymmetric $IV$s may be attributed to two dissimilar metal-molecule bonds rather than to a molecule chemisorbed to a single electrode only. In the latter case the electronic coupling to the other electrode would be characterized by vacuum tunneling, leading to highly stretching-dependent $IV$ characteristics.\\

We have analyzed data from hundreds of $IV$ measurements on MCBJs exposed to benzenedithiol and benzenediamine and identified series of reproducible characteristics. Figure~6 presents typical $IV$ characteristics obtained on several junctions at room temperature and at 77~K together with their respective numerical $dI/dV$ spectra.
\begin{figure}
\label{fig_IVs}
\centering
\includegraphics{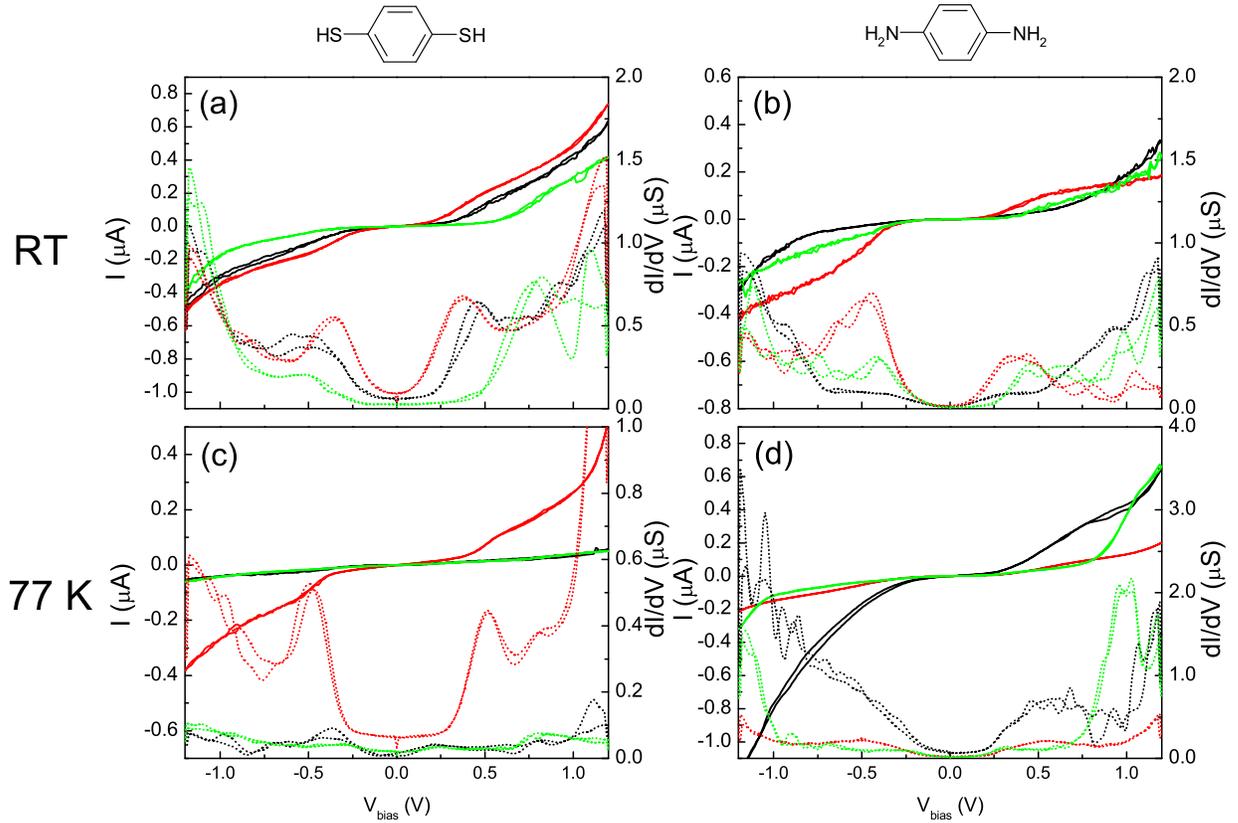}
\caption{Typical $IV$ characteristics (solid lines) of MCBJs exposed to benzenedithiol and benzenediamine. The differential conductance $dI/dV$ (dotted lines) was calculated by numerically smoothing and differentiating the measurement data. The left side displays $IV$ characteristics that remained stable during the stretching of junctions containing BDT at (a) room temperature and (c) 77~K. The right graphs contain typical $IV$ traces of MCBJs containing BDA at (b) room temperature and (d) 77~K.}
\end{figure}
At room temperature, junctions containing benzenedithiol repeatedly displayed conductance gaps at low bias. Three examples of such characteristics are presented in figure 6(a). In this set, the black curve represents a junction with a conductance gap of $2\Delta_c=1.0$~V, framed by two peaks at -0.5 and +0.5~V in the numerical $dI/dV$ spectrum. This gap size is close to the average value in our measurements, which was around 0.9~V. However, $\Delta_c$ showed considerable variation between different $IV$ series, ranging from 0.3~V to 0.8~V. Furthermore, the junction characteristics were not always symmetric. In the graph, this variation is exemplified by the green and the red curves.\\
The non-linear $IV$ characteristics observed at 77~K differ from the ones at room temperature. Measurements typically yielded smaller conductance gaps on the order of $2\Delta_c=0.6$~V, similar to the black and green characteristics in figure~6(c). Occasionally, larger gaps and different current levels could be observed. With two $dI/dV$ peaks located around $\pm$0.5~V the red $IV$ curve in the graph illustrates this variation.\\

The BDT characteristics presented in figure~6 are markedly different from the ones reported by Reed~\etal~\cite{reed97}. In our measurements, the differential conductance beyond the first peak in $dI/dV$ is by one order of magnitude larger. Futhermore, the gap sizes in figure~6(a) and (c) are by a factor of 2.5 smaller than the ones in reference~\cite{reed97}. Our room-temperature results are, however, in good agreement with more recent data of L{\"o}rtscher~\etal~\cite{loertscher07}. At 250~K, these authors reported reproducible $IV$ characteristics with an average $2\Delta_c$ on the order of 0.9~V. In the same study, they obtained a conductance gap of about 0.6~V at 77~K, again in good agreement with the average value in our measurements. However, the wide gap indicated in figure~6(c) differs from the results in reference~\cite{loertscher07}.\\
The presence of varying $IV$ characteristics in single-molecule measurements may be related to changes in electronic coupling or band offset~\cite{troisi06,xue03}. Computational transport studies on BDT junctions have yielded a variation in the position of the electronic levels of the molecule by more than 1~eV~\cite{li07_1,xue03,stadler05,arnold07}. Furthermore, it has been suggested that single-molecule transport is influenced by the stretching of the molecule~\cite{romaner06}, the non-equilibrium situation established by finite bias voltages~\cite{arnold07}, and the adsorption motif of the thiol~\cite{li07_1}. However, the quantitative agreement between theoretical and experimental data is generally too limited for a detailed comparison.\\

Interestingly, the junctions formed in our experiments appear to be more stable than the ones reported earlier. L{\"o}rtscher~\etal~\cite{loertscher07} were only able to establish stable metal-BDT-metal junctions at temperatures of 250~K and below. In our experiments, reproducible characteristics could even be measured at room temperature. Furthermore, the junction currents we observed were by a factor of more than 10 larger. Whilst the similarity in conductance gaps suggests a comparable band lineup of the BDT, the current and stability data support a larger metal-molecule coupling compared to the experiments of L{\"o}rtscher~\etal~\cite{loertscher07}. Regarding the variability of the conductance observed in the histogram measurements on BDT, the signature of many possible junction configurations in $IV$ measurements is not surprising.\\

Measurements on benzenediamine at room temperature revealed conductance gaps on the order of 0.9~V. Typical $IV$ curves are presented in figure~6 (b) and (d). The gaps show a spread similar to the one in BDT junctions. In contrast to BDT, the gap sizes for BDA at 77~K are not systematically smaller than the ones at room temperature.\\
To date, simulations of the electrical properties of benzenediamine junctions are scarce. In a recent theoretical study motivated by the STM-based measurements in reference~\cite{venkataraman06}, Quek~\etal~\cite{quek07} calculated the transmission spectrum of various configurations of BDA between gold electrodes. Their simulations yield an energy difference of 3~eV between the gold Fermi level and the highest occupied molecular orbital (HOMO) of the BDA. In a second study of BDA in gold break junctions, the first molecular level was located at about 1.5~eV from the Fermi energy~\cite{li07_1}. The observation of a conductance gap around 0.9~V supports a very different band lineup in our experiments. However, a direct comparison with the results by Venkataraman~\etal~\cite{venkataraman06} and with calculations should be made with care, as histograms are recorded at low bias. Therefore, they may probe other configurations than the ones that are stable over the wide voltage range required for taking $IV$ curves. Moreover, the position of the energy level in the calculations may not be interpreted directly in terms of the gap in $IV$ as levels can shift under the influence of the applied bias.

\section{Conclusions}
We have studied the electronic properties of single prototypical molecules using lithographic MCBJs in vacuum. Hexane and benzene with thiol and amino anchoring groups were applied using a self-assembly scheme before breaking.\\
Histogram measurements of HDT with different parameters suggest that the resulting concentration of molecules in the junction is low and governed by surface diffusion. Using fusing depths of 20~$G_0$ at room temperature and 3~$G_0$ at 77~K, it was possible to reliably measure single-molecule conductance.\\
In contrast to recent STM-based experiments in solution~\cite{venkataraman06}, the exchange of thiol for amino anchoring groups did not reduce the spread in molecular conductance, neither for hexane nor for benzene cores. Supposedly, the slow breaking and the absence of a surrounding solvent are detrimental for the measurement of conductance histograms. Recent theoretical and experimental data suggest that the amine bond breaks fast under applied stress. Thiols, in contrast, are able to maintain contact upon stretching of the junctions. At low surface coverage and in the absence of bond-reformation, as in MCBJ experiments in vacuum, the lifetime of metal-molecule bond may be the determining factor in conductance measurements. In the case of short-lived junctions and slow breaking, histograms will then be dominated by a uniform background from vacuum tunnelling.\\
Nevertheless, junctions of benzenedithiol and benzenediamine displayed series of reproducible, non-linear $IV$ characteristics during stepwise bending. In contrast to recent BDT measurements by L{\"o}rtscher~\etal~\cite{loertscher07}, which were limited to a maximum temperature of 250~K, we were able to measure stable $IV$s at 300~K. The observed conductance gaps of BDT-junctions were largely in agreement with their results. However, a larger spread in gap sizes and a difference in junction currents indicate the probing of several metal-molecule-metal configurations. The obtained $IV$ characteristics of BDA-junctions at room temperature and at 77~K also displayed some variation, but no systematic difference in conductance gaps at room temperature and 77~K.\\
Our experiments have implications for single-molecule measurements in vacuum. When using lithographic MCBJs the experimental parameters are of large importance to ensure the contacting of single molecules. Furthermore, histogram measurements may be limited by the low surface concentration of molecules and the lifetime of the molecular bond. Similar to histograms, $IV$ measurements can exhibit considerable variation due to the probing of different junction geometries. These results may serve as a guide for future experimental procedures aimed at low-temperature single-molecule characterization.

\ack
The authors acknowledge Ruud van Egmond, Raymond Schouten and Bert Crama for expert technical assistance. Furthermore, they are grateful to Shintaro Fujii and Masamichi Fujihira for providing the benzenedithiol. C.A.M. would like to thank Fedinand Evers, Dago de Leeuw, Sepas Setayesh, and Paul van Hal for helpful advice. All MCBJ devices were fabricated in the Delft nanofacility. This research was carried out with financial support from the Dutch Foundation for Fundamental Research on Matter (FOM).

\section*{References}
\bibliographystyle{unsrt}
\bibliography{martin_njp}

\begin{thebibliography}{10}

\bibitem{reed97}
Reed~M A, Zhou C, Muller~C J, Burgin~T P, and Tour~J M.
\newblock Conductance of a molecular junction.
\newblock {\em Science}, 278:252--4, 1997.

\bibitem{ruitenbeek96}
Van Ruitenbeek~J M, Alvarez A, Pi{\~n}eyro I, Grahmann C, Joyez P, Devoret~M H,
  Est{\'{e}}ve D, and Urbina C.
\newblock Adjustable nanofabricated atomic size contacts.
\newblock {\em Rev. Sci. Instrum.}, 67(1):108--11, 1996.

\bibitem{park99}
Park H, Lim A~K L, Alivisatos~A P, Park J, and McEuen~P L.
\newblock Fabrication of metallic electrodes with nanometer separation by
  electromigration.
\newblock {\em Appl. Phys. Lett.}, 75(2):301--3, 1999.

\bibitem{li98_2}
Li~C Z, Sha H, and Tao~N J.
\newblock Adsorbate effect on conductance quantization in metallic nanowires.
\newblock {\em Phys. Rev. B}, 58(11):6775--8, 1998.

\bibitem{xu03}
Xu~B and Tao~N J.
\newblock Measurement of single-molecule resistance by repeated formation of
  molecular junctions.
\newblock {\em Science}, 301:1221--3, 2003.

\bibitem{haiss03}
Haiss W, Van~Zalinge H, Higgins~S J, Bethell D, Hobenreich H, Schiffrin~D J,
  and Nichols~R J.
\newblock Redox state dependence of single molecule conductivity.
\newblock {\em J. Am. Chem. Soc.}, 125:15294--5, 2003.

\bibitem{donhauser01}
Donhauser~Z J, Mantooth~B A, Kelly~K F, Bumm~L A, Monnell~J D, Stapleton~J J,
  Price Jr.~D W, Rawlett~A M, Allara~D L, Tour~J M, and Weiss~P S.
\newblock Conductance switching in single molecules through conformational
  changes.
\newblock {\em Science}, 292:2303--7, 2001.

\bibitem{cui01}
Cui~X D, Primak A, Zarate X, Tomfohr J, Sankey~O F, Moore~A L, Moore~T A, Gust
  D, Harris G, and Lindsay~S M.
\newblock Reproducible measurement of single-molecule conductivity.
\newblock {\em Science}, 294:571--4, 2001.

\bibitem{vrouwe05}
Vrouwe S~A G, Van der Giessen~E, Van der Molen S~J, Dulic D, Trouwborst~M L,
  and Van Wees~B J.
\newblock Mechanics of lithographically defined break junctions.
\newblock {\em Phys. Rev. B}, 71:035313, 2005.

\bibitem{gonzalez06}
Gonz{\'a}lez~M T, Wu~S, Huber R, Van der Molen S~J, Sch{\"o}nenberger C, and
  Calame M.
\newblock Electrical conductance of molecular junctions by a robust statistical
  analysis.
\newblock {\em Nano Lett.}, 6(10):2238--42, 2006.

\bibitem{loertscher07}
L{\"o}rtscher E, Weber~H B, and Riel H.
\newblock Statistical approach to investigating transport through single
  molecules.
\newblock {\em Phys. Rev. Lett.}, 98:176807, 2007.

\bibitem{tsutsui08}
Tsutsui M, Shoji K, Tanigucho M, and Kawai T.
\newblock Formation and self-breaking mechanism of stable atom-sized junctions.
\newblock {\em Nano Lett.}, 8(1):345--9, 2008.

\bibitem{huang07}
Huang Z, Chen F, Bennett~P A, and Tao N.
\newblock Single molecule junctions formed via {Au}-thiol contact: Stability
  and breakdown mechanism.
\newblock {\em J. Am. Chem. Soc.}, 129:13225--31, 2007.

\bibitem{park07}
Park~Y S, Whalley~A C, Kamenetska M, Steigerwald~M L, Hybertsen~M S, Nuckolls
  C, and Venkataraman L.
\newblock Contact chemistry and single-molecule conductance: A comparison of
  phosphines, methyl sulfides, and amines.
\newblock {\em J. Am. Chem. Soc.}, 129:15768--9, 2007.

\bibitem{smit02}
Smit R~H M, Noat Y, Untiedt C, Lang~N D, Van Hemert~M C, and Van Ruitenbeek~J
  M.
\newblock Measurement of the conductance of a hydrogen molecule.
\newblock {\em Nature}, 419:906--9, 2002.

\bibitem{kergueris99}
Kergueris C, Bourgoin J-P, Palacin S, Esteve D, Urbina C, Magoga M, and Joachim
  C.
\newblock Electron transport through a metal-molecule-metal junction.
\newblock {\em Phys. Rev. B}, 59:12505--13, 1999.

\bibitem{reichert02}
Reichert J, Ochs R, Beckmann D, Weber~H B, Mayor M, and Von~L{\"o}hneysen H.
\newblock Driving current through single organic molecules.
\newblock {\em Phys. Rev. Lett.}, 88(17):176804, 2002.

\bibitem{fujihira06}
Fujihira M, Suzuki M, Fujii S, and Nishikawa A.
\newblock Currents through single molecular junction of
  {Au/hexanedithiolate/Au} measured by repeated formation of break junction in
  {STM} under {UHV}: Effects of conformational change in an alkylene chain from
  gauche to trans and binding sites of thiolates on gold.
\newblock {\em Phys. Chem. Chem. Phys.}, 8:3876--84, 2006.

\bibitem{li07_2}
Li~C, Pobelov I, Wandlowski T, Bagrets A, Arnold A, and Evers F.
\newblock Charge transport in single {Au/alkanedithiol/Au} junctions:
  Coordination geometries and conformational degrees of freedom.
\newblock {\em J. Am. Chem. Soc.}, 130(1):318--26, 2008.

\bibitem{tour95}
Tour~J M, Jones~II L, Pearson~D L, Lamba J~J S, Burgin~T B, Whitesides~G M,
  Allara~D L, Parikh~A N, and Atre~S V.
\newblock Self-assembled monolayers and multilayers of conjugated thiols,
  {$\alpha,\omega$}-dithiols, and thioacetyl-containing adsorbates.
  understanding attachments between potential molecular wires and gold
  surfaces.
\newblock {\em J. Am. Chem. Soc.}, 117:9529--34, 1995.

\bibitem{love05}
Love~J C, Estroff~L A, Kriebel~J K, Nuzzo~R G, and Whitesides~G M.
\newblock Self-assembled monolayers of thiolates on metals as a form of
  nanotechnology.
\newblock {\em Chem. Rev.}, 105:1103--69, 2005.

\bibitem{losic01}
Losic D, Shapter~J G, and Gooding~J J.
\newblock Influence of surface topography on alkanethiol {SAMs} assembled from
  solution and by microcontact printing.
\newblock {\em Langmuir}, 17:3307--16, 2001.

\bibitem{kawasaki00}
Kawasaki M, Sato T, Tanaka T, and Takao K.
\newblock Rapid self-assembly of alkanethiol monolayers on sputter-grown
  {Au(111)}.
\newblock {\em Langmuir}, 16:1719--28, 2000.

\bibitem{basch05}
Basch H, Cohen R, and Ratner~M A.
\newblock Interface geometry and molecular junction conductance: Geometric
  fluctuation and stochastic switching.
\newblock {\em Nano Lett.}, 5(9):1668--75, 2005.

\bibitem{tomfohr04}
Tomfohr J and Sankey~O F.
\newblock Theoretical analysis of electron transport through organic molecules.
\newblock {\em J. Chem. Phys.}, 120(3):1542--54, 2004.

\bibitem{muller06}
M{\"u}ller K-H.
\newblock Effect of the atomic configuration of gold electrodes on the
  electrical conduction of alkanedithiol molecules.
\newblock {\em Phys. Rev. B}, 73:045403, 2006.

\bibitem{haiss04}
Haiss W, Nichols~R J, Van~Zalinge H, Higgins~S J, Bethell D, and Schiffrin~D J.
\newblock Measurement of single molecule conductivity using the spontaneous
  formation of molecular wires.
\newblock {\em Phys. Chem. Chem. Phys.}, 6:4330--7, 2004.

\bibitem{nishikawa07}
Nishikawa A, Tobita J, Kato Y, Fujii S, Suzuki M, and Fujihira M.
\newblock Accurate determination of multiple sets of single molecular
  conductance of {Au/1,6-hexanedithiol/Au} break junctions by ultra-high
  vacuum-scanning tunneling microscope and analyses of individual
  current-separation curves.
\newblock {\em Nanotechnology}, 18:424005, 2007.

\bibitem{jang06}
Jang S-Y, Reddy P, Majumdar A, and Segalman~R A.
\newblock Interpretation of stochastic events in single molecule conductance
  measurements.
\newblock {\em Nano Lett.}, 6(10):2362--7, 2006.

\bibitem{ulrich06}
Ulrich J, Esrail D, Pontius W, Venkataraman L, Millar D, and Doerrer~L H.
\newblock Variability of conductance in molecular junctions.
\newblock {\em J. Phys. Chem. B}, 110:2462--6, 2006.

\bibitem{chen06}
Chen F, Li~X, Hihath J, Huang Z, and Tao N.
\newblock Effect of anchoring groups on single-molecule conductance:
  Comparative study of thiol-, amine-, and carboxylic-acid-terminated
  molecules.
\newblock {\em J. Am. Chem. Soc.}, 128:15874--81, 2006.

\bibitem{tsutsui06}
Tsutsui M, Teramae Y, Kurokawa S, and Sakai A.
\newblock High-conductance states of single benzenedithiol molecules.
\newblock {\em Appl. Phys. Lett.}, 89:163111, 2006.

\bibitem{he05}
He~J, Sankey O, Lee M, Tao N, Li~X, and Lindsay S.
\newblock Measuring single molecule conductance with break junctions.
\newblock {\em Faraday Discuss.}, 131:145–--54, 2005.

\bibitem{venkataraman06}
Venkataraman L, Klare~J E, Tam~I W, Hybertsen~M S, and Steigerwald~M L.
\newblock Single-molecule circuits with well-defined molecular conductance.
\newblock {\em Nano Lett.}, 6(3):458--62, 2006.

\bibitem{xu93}
Xu~C, Sun L, Kepley~L J, and Crooks~R M.
\newblock Molecular interactions between organized, surface-confined monolayers
  and vapor-phase probe molecules. 6. in-situ {FTIR} external reflectance
  spectroscopy of monolayer adsorption and reaction chemistry.
\newblock {\em Anal. Chem.}, 65:2102--7, 1993.

\bibitem{brown99}
Brown~L O and Hutchison~J E.
\newblock Controlled growth of gold nanoparticles during ligand exchange.
\newblock {\em J. Am. Chem. Soc.}, 121:882--3, 1999.

\bibitem{kumar03}
Kumar A, Mandal S, Selvakannan~P R, Pasricha R, Mandale~A B, and Sastry M.
\newblock Investigation into the interaction between surface-bound alkylamines
  and gold nanoparticles.
\newblock {\em Langmuir}, 19:6277--82, 2003.

\bibitem{li07_1}
Li~Z and Kosov~D S.
\newblock Nature of well-defined conductance of amine anchored molecular
  junctions.
\newblock {\em Phys. Rev. B}, 76:035415, 2007.

\bibitem{quek07}
Quek~S Y, Venkataraman L, Choi~H J, Louie~S G, Hybertsen~M S, and Neaton~J B.
\newblock Amine-gold linked single-molecule junctions: Experiment and theory.
\newblock {\em Nano Lett.}, 7(11):3477--82, 2007.

\bibitem{ron98}
Ron H, Matlis S, and Rubinstein I.
\newblock Self-assembled monolayers on oxidised metals. 2. gold surface
  oxidative pretreatment, monolayer properties, and depression formation.
\newblock {\em Langmuir}, 14:1116--21, 1998.

\bibitem{ron94}
Ron H and Rubinstein I.
\newblock Alkanethiol monolayers on preoxidised gold. encapsulation of gold
  oxide under an organic monolayer.
\newblock {\em Langmuir}, 10:4566--73, 1994.

\bibitem{agrait03}
Agra{\"\i}t N, Levy~Yeyati A, and Van Ruitenbeek~J M.
\newblock Quantum properties of atomic-sized conductors.
\newblock {\em Physics Reports}, 377:81--279, 2003.

\bibitem{krans93}
Krans~J M, Muller~C J, Yanson~I K, Govaert T~C M, Hesper R, and Van
  Ruitenbeek~J M.
\newblock One-atom point contacts.
\newblock {\em Phys. Rev. B}, 48(19):14721--4, 1993.

\bibitem{yanson98}
Yanson~A I, Rubio~Bollinger G, Van den Brom H~E, Agra{\"\i}t N, and Van
  Ruitenbeek~J M.
\newblock Formation and manipulation of a metallic wire of single gold atoms.
\newblock {\em Nature}, 395:783--5, 1998.

\bibitem{rubio96}
Rubio G, Agra{\"\i}t N, and Vieira S.
\newblock Atomic-sized metallic contacts: Mechanical properties and electronic
  transport.
\newblock {\em Phys. Rev. Lett.}, 76(13):2302--5, 1996.

\bibitem{xu03_2}
Xu~B, Xiao X, and Tao~N J.
\newblock Measurements of single-molecule electromechanical properties.
\newblock {\em J. Am. Chem. Soc.}, 125:16164--5, 2003.

\bibitem{xiao04}
Xiao X, Xu~B, and Tao~N J.
\newblock Measurement of single molecule conductance: Benzenedithiol and
  benzenedimethanethiol.
\newblock {\em Nano Lett.}, 4(2):267--71, 2004.

\bibitem{troisi06}
Troisi A and Ratner~M A.
\newblock Molecular signatures in the transport properties of molecular wire
  junctions: What makes a junction "molecular"?
\newblock {\em small}, 2(2):172--81, 2006.

\bibitem{xue03}
Xue Y and Ratner~M A.
\newblock Microscopic study of electrical transport through individual
  molecules with metallic contacts. i. band lineup, voltage drop, and
  high-field transport.
\newblock {\em Phys. Rev. B}, 68:115406, 2003.

\bibitem{stadler05}
Stadler R, Thygesen~K S, and Jakobsen~K W.
\newblock An ab inito study of electron transport through nitrobenzene: the
  influence of leads and contacts.
\newblock {\em Nanotechnology}, 16:S155--60, 2005.

\bibitem{arnold07}
Arnold A, Weigend F, and Evers F.
\newblock Quantum chemistry calculations for molecules coupled to reservoirs:
  Formalism, implementation, and application to benzenedithiol.
\newblock {\em J. Chem. Phys.}, 126:174101, 2007.

\bibitem{romaner06}
Romaner L, Heimel G, Gruber M, Br{\'e}das J-L, and Zojer E.
\newblock Stretching and breaking of a molecular junction.
\newblock {\em small}, 2(12):1468--75, 2006.

\end{thebibliography}
\end{document}